# Reflected continuously tunable acoustic metasurface with rotatable space coiling-up structure

Haozhen Zou[1], Yanlong Xu[2], Pan Li[1] and Pai Peng[1],*

[1]School of Mathematics and Physics, China University of Geosciences, Wuhan 430074, China

[2]School of Aeronautics, Northwestern Polytechnical University, Xi'an 710072, Shaanxi, China

## Abstract

In this paper, we propose a continuously tunable acoustic metasurface composed of identical anisotropic resonant units, each of which contains a rigid pedestal and a rotatable inclusion with space coiling-up structure. The metasurface can manipulate the reflected phase by adjusting the rotational angle of inclusion. The theoretical analysis shows that the polarization-dependent phase change can be induced by the even-order standing wave modes inside inclusion. By utilizing the rotatable inclusion, we design a tunable acoustic carpet cloaking device, which works with a wide range for incident angle. When incident waves come from different directions, the cloaking effect can be obtained by arrange the rotational angle of each inclusion.

*Corresponding author: paipeng@cug.edu.cn

Acoustic metasurfaces (AMs)[1] that offer phase modulation have attracted a lot of interest in recent years. Generally, AMs can be constructed using one layer of units, which can provide large phase delays relative to the matrix around the working frequency. Several kinds of units have been wildly used and systematically studied in AMs, such as the Helmholtz resonators[2], the three-component resonators[3, 4], membranes-type structure[5] and space coiling-up structure[6]. These classical AMs (CAMs)[7-13] straightly composed of these acoustic units mentioned above are with fixed geometries and can only achieve a single special functionality set in advance. To overcome the restriction, researchers put forward a new structure, the tunable acoustic metasurfaces (TAMs)[14, 15], which can flexibly adapt to the different incident frequencies and incident angles through modulating its tunable units. Sheng-Dong Zhao et.al proposed a continuously TAMs by screwing the nut[14], which can adjust the effective spatial length of the space coiling-up structure, to modulate the transmitted and reflected wavefront. Pan Li et.al proposed a broad-band reflected TAMs by rotating the elliptical rotor of the modified three-component unit[15], which can change the polarization direction of the dipolar resonance, to modulate the reflected wavefront. Although so far, there is more and more attention being paid to the research on TAMs, the achievement is still a little.

In this paper, we design a continuously TAM by utilizing anisotropic unit with space-coiling structure. The unit is composed of a rigid pedestal and a rotatable space-coiling inclusion. The reflected phase is controlled by the rotational angle of inclusion. A TAM is proposed with the identical units, and the TAM properties depend on the rotational angle profile. We further demonstrate a wide-angle carpet clacking device[16] by using the TAM.

As shown in Fig.1(a), the TAM we designed is constructed by a series of identical anisotropic resonant units. These units are periodically arranged along the x-direction and placed in an air background, whose material parameters are $\rho = 1.29 \text{kg/m}^3$ for mass density and $c = 340 \text{m/s}$ for wave velocity. Each anisotropic unit can be divided into two parts. As shown in Fig.1(b), the first part is a rigid pedestal with width of $l_y$ and length of $h = 1.1718 l_y$. The second part is a rigid rotatable inclusion with space coiling-up structure. The width and length of the inclusion are $a_x = 0.625 l_y$ and $a_y = 0.625 l_y$, respectively. There are two symmetric openings in the middle of top and bottom surfaces. The total spatial length from the top opening to the bottom opening inside the inclusion is about $L_0 \approx 8.4687 l_y$. The inclusion could rotate around its center with arbitrary rotated angle $\beta$. The distance from the center to the pedestal's sides and bottom are $r_y = 0.4844 l_y$ and $r_x = 0.4844 l_y$, respectively. The other detail parameters could be found in Fig.1 caption.

We let plane pressure waves (with wavelength $\lambda_0$) normally incident from the top and calculates the reflected phase. The phase spectra for $\beta = 0°$ and $\beta = 90°$ are plotted in Fig.2(a) by the black dash and red solid lines, respectively. In the case of $\beta = 0°$, the phase has a large change about $2\pi$ around two special wavelengths of $l_y / \lambda_1 = 0.0588$ and $l_y / \lambda_2 = 0.1178$. In the case of $\beta = 90°$, the large phase change are only found around the $\lambda_1$, but there is almost no phase change for the $\lambda_2$. That means a large phase change could be dependent on the polarization of the rotatable inclusion.

Now we focus on the two special wavelengths of $\lambda_1$ and $\lambda_2$. The pressure fields of the unit at $\lambda_1$ and $\lambda_2$ are plotted in Fig. 2(b-e). We can see that they are exhibit as the first-order and second-order standing wave modes inside the inclusion. The wavelengths of $\lambda_1$ and $\lambda_2$ indeed satisfy to the required wavelengths of the first-order ($\lambda_1 \approx 2L_0$) and second-order ($\lambda_2 \approx L_0$) standing wave. In addition, the first-order of standing wave mode seems insensitive to the rotational angle $\beta$, but the existence of the second-order of standing wave mode depends on the polarization direction of the inclusion. As shown in Fig. 2(e), the pressure field is nearly zero when the inclusion is horizontally placed. Because here the second-order standing wave mode does not seem to be excited, there is no large phase change.

In order to qualitatively explain the dependence on polarized direction for the second standing wave mode, we theoretically study the pressure properties inside the inclusion. As shown in Fig. 2(f) and (g), the inclusion could be modeled as a narrow straight waveguide with width $w_y \ll \lambda_0$ and length $L_0$[6]. The pressure on the input and output surfaces outside the waveguide are $p_1 = p_0 e^{i\varphi_1}$ and $p_2 = p_0 e^{i\varphi_2}$, respectively. Here $p_0$ is the amplitude of the incident waves, and $\varphi_1$ and $\varphi_2$ are the initial phases related to the positions of input and output surface, which are actually dependent on the rotational angle $\beta$. The pressure wave equation inside the waveguide can be expressed as:

$$p(r) = p_+ e^{ikr} + p_- e^{-ikr}, \tag{1}$$

where $p_+$ and $p_-$ are unknown amplitudes, $k = 2\pi/\lambda$ is the incident wave number and $r$ is the spatial coordinate in the waveguide. The boundary conditions on the input and output surfaces are

$$\begin{cases} p(r=0) = \alpha p_0 e^{i\varphi_1} \\ p(r=L_0) = \alpha p_0 e^{i\varphi_2} \end{cases}, \qquad (2)$$

where $\alpha$ is a small coefficient related to the geometry of the waveguide. By applying Eq. (1) to Eqs. (2), we can obtain the equations about $p_+$ and $p_-$ as:

$$\begin{cases} p_+ + p_- = \alpha p_0 e^{i\varphi_1} \\ p_+ e^{ikl} + p_- e^{-ikl} = \alpha p_0 e^{i\varphi_2} \end{cases}. \qquad (3)$$

After solving the Eqs. (3), the Eq. (1) can be expressed as:

$$p(r) = \alpha p_0 \left( \frac{e^{i\varphi_1} e^{-ikL_0} - e^{i\varphi_2}}{e^{-ikL_0} - e^{ikL_0}} e^{ikr} + \frac{e^{i\varphi_1} e^{ikL_0} - e^{i\varphi_2}}{e^{ikL_0} - e^{-ikL_0}} e^{-ikr} \right). \qquad (4)$$

For the special cases that the n-order standing wave modes, the length of waveguide $L_0$ should equal to an integer ($n$) multiple of the half-wavelength $\lambda_0/2$. We notice that actually they are not always strictly equally ($L_0 \approx n\lambda/2$). Here we set $L_0 = n\lambda/2 + l$ with $l$ an infinitely small value and employ the Taylor expansion, and then we get $e^{ikL_0} \approx (-1)^n (1+ikl)$ (and $e^{-ikL_0} \approx (-1)^n (1-ikl)$) after ignoring the high order terms. By employing this approximation, Eq. (4) can be reduced to:

$$p(r) = \alpha p_0 \frac{\lambda_0}{l} \frac{(-1)^n e^{i\varphi_2} - e^{i\varphi_1}}{2\pi} \sin\left(\frac{n\pi}{L_0} r\right). \qquad (5)$$

When n is an even number, Eq. (5) changes to:

$$p(r) = \alpha p_0 \frac{\lambda_0}{l} \frac{e^{i\varphi_2} - e^{i\varphi_1}}{2\pi} \sin\left(\frac{n\pi}{L_0} r\right). \qquad (6)$$

The term $e^{i\varphi_2} - e^{i\varphi_1}$ shows that the pressure $p(r)$ depends on the phase difference between $\varphi_1$ and $\varphi_2$, which can be roughly regard as $\Delta\varphi = \varphi_2 - \varphi_1 = k_0 a_x \cos\beta$. In particular, when the waveguide is horizontally placed, we get $\beta = 90º$ and $\Delta\varphi = 0$.

Thus Eq. (6) will reduce to trivial zero, which agrees with the simulated pressure field as shown in Fig.2(e). In contrast, if the waveguide is vertically placed ($\beta = 0$), we have $\varphi_1 \neq \varphi_2$ and the pressure $p(r)$ expresses as a classical solution for standing wave. The corresponding (n=2) simulated result is shown in Fig. 2(d), which agree with the theoretical prediction. When n is an old number, Eq. (5) changes to:

$$p(r) = -\alpha p_0 \frac{\lambda_0}{l} \frac{e^{i\varphi_2} + e^{i\varphi_1}}{2\pi} \sin\left(\frac{n\pi}{L_0} r\right) \tag{7}$$

Be different to Eq. (6), here the pressure has not trivial solution for neither $\beta = 0$ nor $\beta = 90°$. When the waveguide is vertically or horizontally placed, the pressure fields are always exhibited as classical standing wave modes, as shown in Fig.2(b) and Fig.2(c), respectively.

As the existence of the second-order of standing wave mode is dependent on the polarization of inclusion, we can use the polarization to control the reflected phase. When the rotational angle $\beta$ continuously increases from $\beta = 0°$ to $\beta = 90°$, the second-order of standing wave mode would gradually vanish, and the reflected phase is excepted to change $2\pi$. We choose $\lambda_1 = 0.11859 l_y$ as the working wavelength and study the relationship between the reflected phase $\varphi$ and the rotational angle $\beta$. The result is shown Fig.3(a), where the phase change $\varphi$ can gradually decrease with the increasing of rotational angle $\beta$. The phase change is discretized into ten elements with a step of $\pi / 5$, as marked in Fig.3(a) by the red dots. The corresponding pressure fields are shown in Fig.3(b). We can manipulate the reflected phase by adjusting the rotational angle $\beta$.

Now we can make tunable carpet cloaking devices by using TAMs with the tunable units shown in Fig. 3(b). Be different with the previous devices[10, 11] that made by classical space-coiling units with fixed geometries, we purpose the tunable device is with a wide range of incident angle. As shown in Fig.3(c), a rigid bump in the form of a triangular prism is placed on the bottom, where $x = 0$ is the symmetry axis of the triangle. The bump has a height of $w(0) = 13.681 l_y$ and base angles of $\alpha = 20°$. When an acoustic Gaussian beam normally incident to the horizontal wall from top, the scatting waves (as shown in Fig.3(c)) will carry the information about the bump. We suppose to "cancel" the scatting by sticking a layer of TAM on the prism surface, as shown in Fig.3(d). One TAM is constructed by 80 identical unit cells shown in Fig.1(b).

According to the generalized Snell's law[17], the additional wave vector (phase gradient) along the AMs direction can be defined as $\xi = d\varphi / dl$, where $d\varphi$ is the discretized phase step and $dl$ is the distance between two neighboring units. The properties of reflected waves can be obtained in the form

$$\begin{cases} k \sin(\theta_r + \alpha) = \xi_{\text{left}} + k \sin(\theta_i - \alpha) \\ k \sin(\theta_r - \alpha) = \xi_{\text{right}} + k \sin(\theta_i + \alpha) \end{cases}, \quad (8)$$

where $\theta_i$ and $\theta_r$ are the incident and reflected angle of the Gaussian beam, respectively. In order to cancel the scattering, we need $\theta_r = \theta_i$. In the case of normal incidence $\theta_i = 0°$, we can obtain the phase profile from Eq. (8) as $\varphi(x) = \varphi(0) - 2k \tan(\alpha)|x|$. Base on the phase profile $\varphi(x)$ and the relationship between $\varphi$ and $\beta$ (as shown in Fig.3(a)), we can obtain the rotational angle profile $\beta(x)$, which is plotted in Fig,3(d) by the black line. The reflected pressure field is displayed in Fig.3(d), where plane reflected waves are clearly observed. The reflected field is essentially the same as the field reflected by the wall without the bump. When the incident angle is changed to $\theta_i = 60°$, the obtained phase profile is: $\varphi(x) = \varphi(0) - k\left[\sin(\theta + \alpha) - \sin(\theta - \alpha)\right]|x|/\cos(\alpha)$. The rotational angle profile is obtained in a similar way and plotted in Fig.3(f) by the black line. The total pressure field is displayed in Fig.3(f) which agrees with the field without the prism as shown in Fig.3(e) where the Gaussian beam is just reflected by a horizontal smooth wall. Figure 3(h) shows the results without the TAMs, as a reference, we can see that the TAM indeed improves the effect of cloaking. As can be seen from Fig.3(d) and Fig.3(f), the proposed TAM can keep the cloaking effect with different incident angle by adjusting the rotational angle profile. In contrast, if the cloaking device is made by CAM with fixed geometry, the cloaking effect will fail when the incident angle is changed. For example, if we consider CAMs have the same structures to the TAMs shown in Fig. 3(d), the cloaking effect could be successfully obtained for $\theta_i = 0$ but failed for $\theta_i = 60°$ as shown in Fig. 3(g). We need to replace new CAMs for different incident angles. Unlike the CAMs, the proposed TAM can work well with different incident angles, showing good performance in terms of the carpet cloaking.

In conclusion, the proposed TAM is built by a series of identical anisotropic resonant units. We can modulate the reflected phase through rotating the inclusion which could change the initial phase difference between the two ends of the inclusion to adjust the extra phase to the reflected waves. As a performance, we concentrate on acoustic carpet clacking and show the reflected field and total field in different cases (the bump is covered by TAMs, CAMs, and nothing) and with different incident angle. Form the results, the TAMs have a good performance contrasting to CAMs and have good potential in the application.

## Acknowledgement

This work was supported by the National Natural Science Foundation of China (Grant No: 11604307)

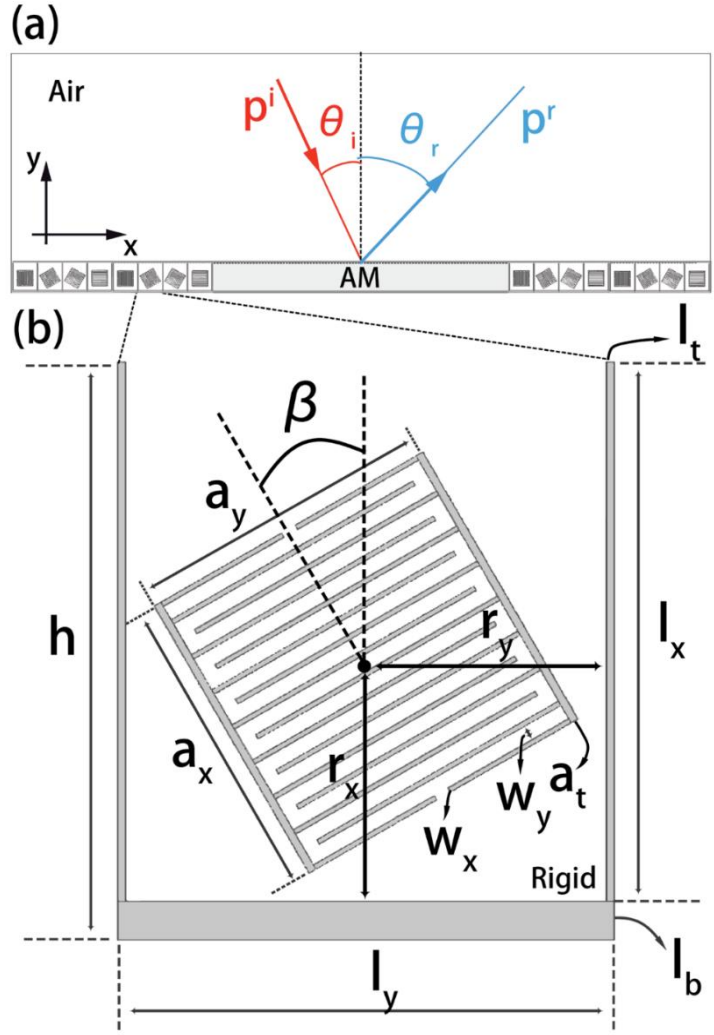

**FIG.1** (a) The schematic diagram of the TAM. (b) The schematic diagram of one unit. For the pedestal, the thickness and length of the left and right partitions are $l_t = 0.0156 l_y$ and $l_x = 1.0937 l_y$, respectively. The thickness and length of the bottom rigid plate are $l_b = 0.0781 l_y$ and $l_y$, respectively. For the inclusion, the thickness of the left and the right rigid plate is $a_t = 0.0156 l_y$. The thickness of each rigid bar inside the inclusion is $w_x = 0.0098 l_y$. The gap widths between each bar and the widths of two openings are equal as $w_y = 0.0312 l_y$. The total spatial length from the top opening to the bottom opening inside the inclusion is about $L_0 = 14(a_x - 2a_t) + 16 w_x = 8.4687 l_y$.

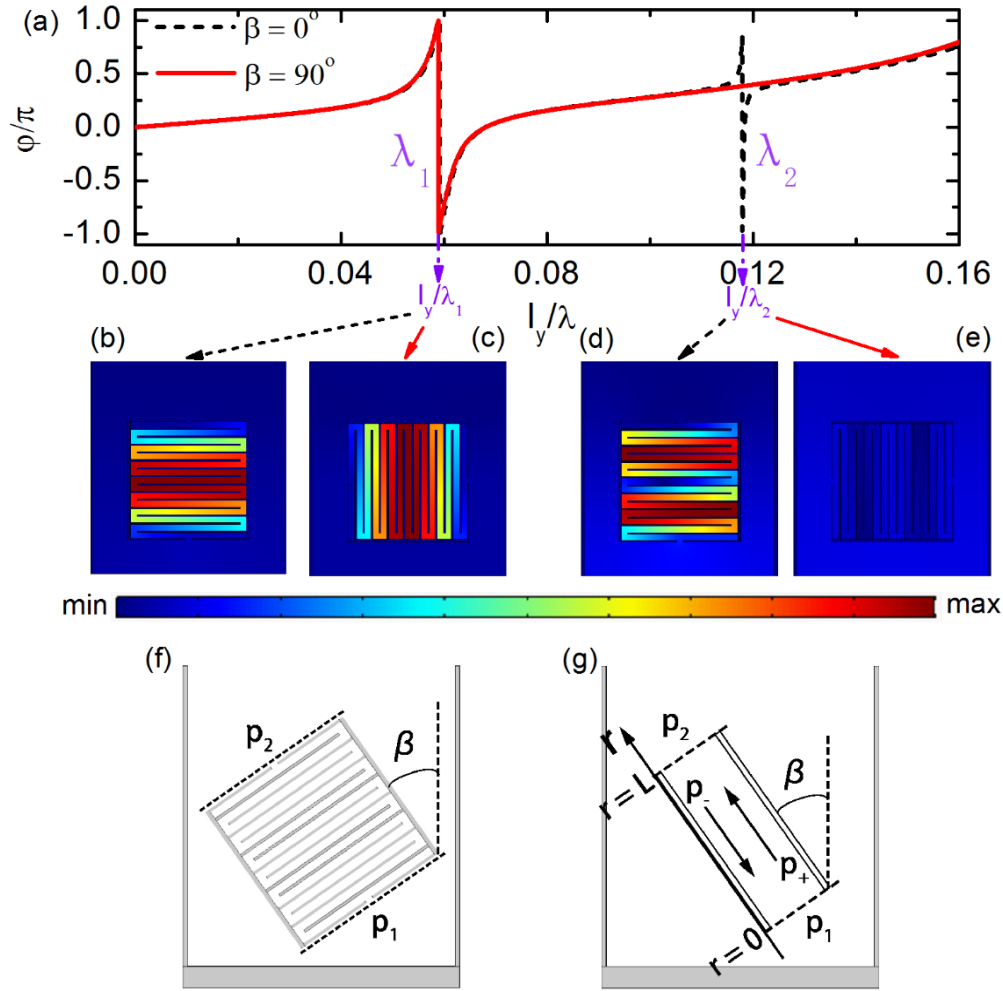

**FIG.2** (a) The black dash and the red solid lines show the reflected phase with $\beta = 0°$ and $\beta = 90°$, respectively. (b), (c), (d) and (e) are the pressure fields in one unit for the cases of $\beta = 0°, \lambda = \lambda_1$, $\beta = 90°, \lambda = \lambda_1$, $\beta = 0°, \lambda = \lambda_2$ and $\beta = 90°, \lambda = \lambda_2$, respectively. (f) The schematic diagram of one unit with arbitrary rotated inclusion. (g) The schematic diagram of waveguide model corresponding to (f).

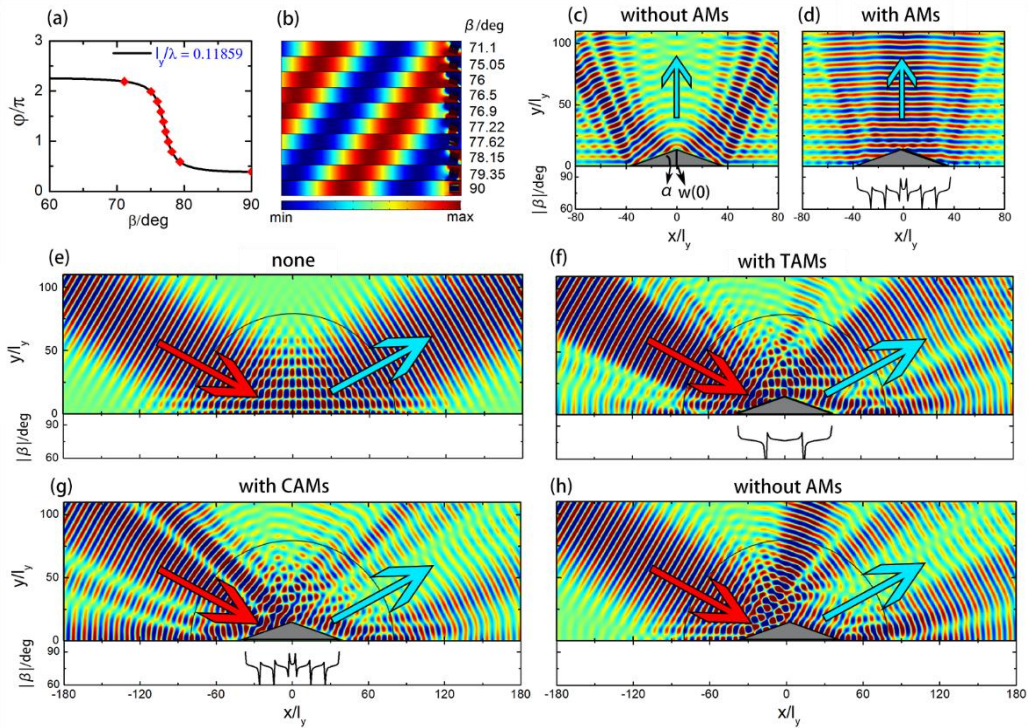

**FIG.3** (a) The reflected phase shift as a function of the rotational angles $\beta$ at the working wavelength $\lambda_2$. (b) The reflected pressure fields of the ten units with exact rotational angles corresponding to those marked in (a) by red circles. (c) and (d) are the reflected pressure fields with and without the designed AM cloak under a normal incident wave, respectively. (e-h) are the total pressure fields with TAMs, without AMs, with CAMs and with nothing under an incident angle $60°$. The red arrows show the direction of incident waves and the blue arrows show the expected direction of the reflected waves.